# The April 1, 2471 b.c. eclipse and the end of the 4th Egyptian dynasty

**Giulio Magli**[1]

*[1]Department of Mathematics, Politecnico di Milano, Italy.*
*Giulio.Magli@polimi.it*

On April 1, 2471 b.C. an impressive, unpredictable phenomenon occurred over the Delta of the Nile: a total solar eclipse, with totality band almost centered on the sacred city of Buto, and the "capital" Memphis very close (>95%) to totality. This date is compatible with existing chronologies for the reign of Pharaoh Shepseskaf, who adopted a clamorous symbolic break with respect to the tradition of "solarized" kings started by Khufu. Indeed his tomb was not built in view from Heliopolis and was not a pyramid, but a kind of unique monument resembling the symbolic shrine at Buto. The aim of the present paper is to investigate in a systematic way the possibility that the origin of this historical passage, which marks the end of the 4th dynasty, can be identified precisely in the 2471 b.C. eclipse, therefore furnishing a new astronomical anchor for the chronology of the Old Kingdom.

## 1. INTRODUCTION

As is well known, the division of the history of Egypt into dynasties comes from the Hellenistic historian Manetho. It is not a division along distinct bloodlines, but rather a framework based on discontinuities, or on events that the author perceived as such. In spite of its late origin, this division catches key moments in the history of Egypt, and one of these key moments by all means is that occurred with the last ruler of the 4th dynasty, Shepseskaf, and the subsequent advent of the 5th dynasty. The details of this passage are not clear from the point of view of the royal bloodline, which might have been unaltered, but the differences between the IV and the V dynasty are striking, and are clearly marked by the Shepseskaf reign. This king made groundbreaking choices for his funerary monument, which appear as an explicit refusal of the solar tradition established by his predecessors. The reasons behind such a break have never been satisfactorily explained. It is the aim of the present paper to analyse in full details the possibility that it was a total solar eclipse, occurred over the Delta of the Nile on April 1, 2471 b.C. (all dates in the present paper are in proleptic Julian) to trigger this interesting and fascinating turnaround in the history of ancient Egypt.

## 2. THE SHEPSESKAF BREAKTHROUGH

Shepseskaf was the successor, almost probably the son, of Menkaura, the builder of the third Giza pyramid. His reign is certainly to be located between the end of the $26^{th}$ century and the mid of the $25^{th}$ century b.C. and should have lasted four to seven years.
Information about Shepseskaf are scarce; we have, however, his tomb (identified by Jéquier in 1925). In the Old Kingdom the Pharaoh's tomb was a fundamental symbol of his divine rights as ruler. These monuments are, therefore, also an indirect but substantial source of information on their builders. Further, we have two main epigraphical sources. The first is the (fragmentary) stela of the king's decree concerning the architectural completion and the offering to be made in the Menkaura complex at Giza, found in the Menkaura funerary temple (Reisner 1931). The second is the surviving text in the registers of the royal annals (the Palermo stone), from which we shall start our analysis.
The text refers to the first regnal year (I will use here the translation by Breasted 1906; points of disagreement will be signaled). The king "appears, he who unites Upper and Lower Egypt" and performs a series of ritual acts: he "circumambulates the wall; (presides a) diadem-festival; orders two (statues of?) Wepwawet to be created" (Wepwawet, or Upuaut, depicted as a jackal on a boat, was the God "opener of the ways" of afterlife). The text proceeds with the king "following the gods who unite the two lands" and a word which might mean "suppliers" (for an alternate meaning see Bogdanov 2019). Then, the act of "choosing the location of the pyramid" is mentioned, together with the name given to the monument (from Snefru onward each pyramid received a name, for instance "Khafra is great" or "Menkaura is divine"). The name of the Shepseskaf pyramid was . Breasted translates this "literally" – that is, with the meaning of the object represented in the hieroglyph - as "The Pyramid (called) fountain of Shepseskaf", but the correct meaning is certainly "The Pyramid (called) Shepseskaf is purified". Finally, the Annals say that the king ordered the construction of a "shrine of Upper and Lower Egypt".
Mention of the ceremonies performed in first year is common to the annals, although the "Diadem" one is barely attested; the statement about the two statues is rather unusual, but most of all what is surprising is the explicit mention of the act of choosing the location of the pyramid. As far as I know indeed this  the unique text not only of the Palermo stone but of all the Old Kingdom in which the *act of choosing the place* for the pyramid of a Pharaoh is explicitly mentioned. Why? The reason probably is that the choice in question represents *a complete breakthrough* with respect to the tradition established by all the four predecessors of the king, and furthermore that the *shape* of the monument also represents *a complete breakthrough* with respect to the tradition established by all the five  IV dynasty  predecessors of the king. In addition, it will remain a unique case later on. We shall now discuss these two points in details, starting from the choice of the place. To do this, I need to recall briefly the topography of the IV dynasty pyramids fields (Magli 2010).
The founder of the fourth dynasty, Snefru, built his two pyramids in Dahshur. These two pyramids were probably conceived as a unitary project, forming – when seen from Saqqara - a huge two-mountain hieroglyph , a sign associated with the cult of the deaths since the early dynastic necropolis in Abydos (Belmonte & Magli 2015).  With his son, Khufu, we assist to a solarization of the divine nature of the Pharaoh (Hawass 1993). Khufu will indeed build his pyramid on the Giza plateau, in plain view from Heliopolis (the main theological centre of the cult of Ra), and will make an explicit reference to the sun cult with the spectacular hierophany occurring at Giza at the summer solstice  re-creating the "solarized" version of the double mountain sign, Akhet (Lehner 1985, Shaltout et al. 2007, Magli 2008). The successors of Khufu (Djedefra, Khafra, Menkaura) will all adopt the suffix -Ra in the their name and construct their pyramids in view of Heliopolis as well (at Abu Roash and Giza respectively; Lehner 1985, Jeffreys 1998, Magli 2010,2013). It should be noted that a further pyramid which is almost certainly a IV dynasty project was left at the very first stage of construction in Zawiet el Arian, again in view from Heliopolis. The king who commissioned this project is unknown (sometimes a king called Bikeris in Greek sources is advocated, but there is no mention of him in Egyptian lists).  The only information we have on the owner are some drawings made by the excavator Barsanti who copied crude inscriptions he found on some blocks (the site is unreachable since many years, being inside a military base) (Barsanti & Maspero 1907). In the drawings the cartouche of a king appears, with two symbols still visible inside: one is Ka, the other is a sketchy thing which has been interpreted in various ways but, at least in the author's view, may be -špṣ. Thus the solution for the "mysterious" owner of this "Great Pit" can be that this was a tomb initiated by Shepseskaf in the first month of his reign, but almost immediately abandoned for the reasons we're going to explain.

The place chosen by Shepseskaf is a plateau located to the south of Saqqara, on the way to Dahshur and non inter-visible with Heliopolis. This area was, at that time, virgin soil (the pyramids visible today are of the end of the V and of the VI dynasty).  The symbolic signal is very clear: staying close to the very ancient tradition of the III Dynasty (the Step Pyramid of Djoser) but "midway" also to the "pre-solar" roots of the IV  (the Bent and the Red pyramids of Snefru). Further, as put in evidence some years ago (Magli 2010, 2013) the peculiar shape of the monument (to be discussed below) implies that it  was conceived to "own" the horizon, exactly as the monuments of the king's predecessors in Giza and Dahshur were. In fact, it was deliberately located  in such a way that the line of sight from the Saqqara central field (to fix ideas, say from  the "entrance" area located near the Teti pyramid) frames the king's tomb as a sort of regular baseline for the double-mountains symbol created at the horizon by the two giant pyramids of Snefru (Fig 1).
All in all, although the placement of the pyramids did depend also on practical aspects such as availability of stone (Barta 2005), the case of Shepseskaf is one of the most clear examples of a monument placed exactly where  the king wanted it to be. The idea was to "complete" the landscape of power built by Snefru, establishing in this way his own power and conveying *a message of order and return*.
Let us now analyse the shape, which at a distant sight may resemble a Mastaba, so that it is traditionally called *Mastabat Faraun.* When the monument is approached by walking through the desert from the Saqqara village, it is easily seen that the analogy with a Mastaba is an illusory effect (Fig. 2) . The building is indeed rectangular in plan, but it is enormous and made of extremely regular courses of huge stones. It is today 99.6 metres long and 74.4 metres broad, but it was certainly meant to measure 200x150 cubits including the casing, which was in granite for the lower courses. The sides are very well orientated to the cardinal points (within half of a degree of true north if a precision magnetic compass corrected for magnetic deviation is used, but a measure with a more precise instrument would be certainly worth making). The walls are 18 metres high and slope inward at 70°; their top was raised up at the two lateral ends thanks to two low vertical walls (today barely recognizable). All in all, the body of the building was clearly constructed by a royal architect with the accuracy and the refined techniques employed for (and exclusively for) the fourth dynasty royal pyramids (for a detailed description see Hamilton 2018).

As far as the interior is concerned, I was granted several years ago by the Egyptian authorities (former SCA) of the permission of visiting it (the whole area is closed to the public) and I can testify that it is really a IV dynasty royal monument also inside. As in all Old Kingdom pyramids, the entrance is in the north face, from which a descending corridor enters the funerary apartments. These apartments have nothing to do with those of a standard Mastaba tomb; instead they bear resemblance to the apartments of Menkaura's pyramid, and are made of perfectly joined, huge granite blocks. Scattered pieces of what seems to be the king’s coffin, of a very hard stone, lie in the antechamber (it is difficult to imagine who – and also when, how and why - managed the task of destroying it).

All in all, the unprecedented and unequalled shape of the Shepseskaf tomb is really a mystery. A number of proposed solutions for this mystery is available, most of them being frankly very weak, or even untenable. For instance, it has been proposed that it was a temporary backup for his own tomb while the king was finishing Menkaura’s complex (Verner 2002) – but this blatantly conflicts with the raffinate interiors. Another proposed solution is that the king may have lacked full legitimacy for some reason (maybe ascending the throne through marriage, or overruling a more legitimate heir) and therefore choose a “modest” monument (the same objection holds). The most untenable of all proposals is Quirke (2001) idea that what we see are the remains of a unfinished step pyramid later adjusted as a mastaba (anyone who has visited the monument can confirm that it was built in accordance with a unitary project). More close to the reality, Hassan (1936) has put forward the idea that Shepseskaf may have deliberately chosen to build a original monument to differentiate it from the pyramids, because these were too much associated with the Sun cult, an idea we will explore in details in a while.

In *any* case, which are the reasons that dictated the *specific* choice of such a peculiar shape? There exist several sarcophagi of the Old Kingdom whose lids have raised tops, and in the Giza tomb ST11, belonging to the Vth dynasty overseer Nikauhor, the monument is designed with the hieroglyph (which however also means “tomb”). So, it has been proposed that the Mastabat Faraun resembles a giant sarcophagus. This interpretation however makes little sense: as far as we know pharaoh’s sarcophagi did not have raised tops during the 4th dynasty, while the ceiling of the funerary chamber of Shepseskaf, similar to what can be seen in Menkaura pyramid, is in itself shaped into a curved “coffin” profile.

Another, more reasonable interpretation is that the tomb resembles a specific, sacred building: a *Buto shrine* (Lehner 1999). The form of these archaic edifices (probably made of perishable materials) is known from the corresponding hieroglyph, representing an arched roof building with side poles . Buto was a important cult centre, located in the delta. The Cobra-goddess Wadjet was worshipped there; together with the vulture God Nekhbet of upper Egypt, they formed the “two ladies” , the patrons of the kingship. Scant remains of the ancient settlement at Buto have so far been found, but sample excavations have shown that the site was already inhabited in pre-dynastic times and that it was a fairly important sacred place, being the northern counterpart of Hierakonpolis (interestingly enough, very recent excavations at Buto identified a 6th century b.C. building which, at least according to the archaeologists in charge of the site, might be identified with an “astronomical observatory”).
The idea that the king may have replicated a hieroglyph, rather than an existing building, should not come as a surprise. Indeed the interplay between symbolic hieroglyphs and architecture is typical of the way the Egyptians conceived writing, as magisterially put in evidence by Assmann (2007). According to Assmann the very idea of creation was tied up with writing in the Egyptian's mind; in a sense the pyramid was a “sign” within a “written” sacred landscape. With the IV dynasty the pyramids become “gigantic hieroglyphs”, as Lehner (1999) puts it; first the double-mountain sign with Snefru, and then an icon of glory – Akhet, the Sun between the two mountains – with Khufu (Magli 2016).
All in all, the tomb of the king speaks to us – both in its placement not in view from Heliopolis, and in its shape connected to a sacred place in the Delta – about a strong break operated by the king with respect to the solar cult; this is confirmed by the simple fact that, contrary to all his three predecessors (Djedefra, Khafra, Menkaura), he declined to have a particle ☉-*Ra* in his name. The almost traumatic relevance that this break must have had can be seen also *a posteriori* by the complexity of the process of restoration and renewal initiated by his successor: the founder of the fifth dynasty Userkaf.
Userkaf's relationships with the royal family are yet unclear, as well as the role played in his accession by queen Khentkaues I, owner of the huge funerary monument located close to Menkaura's causeway (Verner 2006). The political programme of return to the solar tradition operated by him is anyhow documented, although through the medium of tales, in the later Westcar Papyrus, where the new generation of kings started with Userkaf is credited of a direct lineage of the Sun God, who - is said - made pregnant the wife of one of his priest.
From the architectural point of view, Userkaf devised a complex way of reconciling all the previous traditions within his funerary project, re-introducing the pyramid as tomb but operating an explicit recall to the Sun and Heliopolis with a second building. The tomb is in the Saqqara central field, thus “taking at distance” the 4th dynasty Snefru tradition and actually staying as close as possible to the Step Pyramid. The interior apartments are made of huge blocks, but the pyramid is modest in dimensions and quality with respect to the IV dynasty projects. The second building – usually called Sun Temple – is in a hitherto virgin area, that of Abu Gorab. It is a new kind of religious ensemble, structured as a pyramid complex (that is, endowed with a Valley building and a causeway) but centered on a rectangular enclosure with a central mound, later substituted by an obelisk, both being clear references to Heliopolis and to the Sun cult. Actually Abu Gorab is the last area of the west bank inter-visible from Heliopolis, before the block of the view created by the outcrop of the Cairo citadel (Jeffreys 1998). The following kings, starting with Sahura, will inaugurate a new Necropolis in nearby Abusir and render these choices a constant of the dynasty (for complete references and details on the topography of the Sun Temples see Magli 2023).
With the founding of the new necropolis, a lineage of kings “sons of Ra” was definitively established and was to endure for three generations. The Shepseskaf break was not without consequences, tough: the structure of the state remained stable, but the members of the royal family ceased to be the unique holders of the highest offices of the state. In the meanwhile, however, the renewed and enhanced role of Ra as “state God” helped to maintain the primacy of the ruler.

## 3. THE SHEPSESKAF ECLIPSE

As Assmann (2001) observed, in the Old Kingdom the religious world was perceived as a global interaction between the community and the gods, with the Sun God in a prominent role. This interaction perpetuated Maat, the Cosmic Order, the king being the mediator and the keeper of such order; by necessity, the funerary monuments of the kings formed an equally ordered landscape. For these reasons, the above described break perpetuated by Shepseskaf – a break visible in the throne name, in the location and shape of its tomb, and also in other details, such as the maniacal attention for the duality of Upper and Lower Egypt in the royal annals, and the name of the tomb which attributes a rather uncommon adjective for a king - must have had a very important, unescapable motivation, and this motivation likely originated in a sudden, unpredictable perturbation of the cosmic order involving the Sun.
Furthermore, this perturbation should have been seen as a omen impossible to keep secret.
Clearly, a total solar eclipse fits well with this description.
Actually, regarding eclipses some mechanism of avoidance must have been in action in ancient Egypt, since there is no known text where an eclipse description can be unambiguously identified. References to eclipses may, however, be present in allusive form; for instance, in the dedication to Tutankhamen of a stela of Huy, overseer of Nubia, we can read "I see the darkness during the daylight (that) you have made, illumine me so that I may see you" (Rowe 1940). Furthermore, as repeatedly noticed in the literature, the frequent and curious image of the winged sun bears an impressive resemblance with the total eclipse fares (Brewer 1991, Sellers 1992) and there are cases in which a tantamount relevance of eclipse omens in Egyptian history is difficult to negate. A clear cut one is the solar eclipse occurred on May 14, 1338 BC, whose path of totality crossed Middle Egypt. The date is very close to the presumable date of foundation of Akhet Aten, the city founded by Akhenaten as the apogee of his monotheistic reform in favour of the solar disk, whose placement – according to the foundational stele - was dictated to the king directly by the God (McMurray 2005, Magli 2013, Belmonte & Lull 2023). So, we are led to search if a total solar eclipse occurred on Lower Egypt in a date compatible with Shepseskaf first year. This eclipse actually exists and occurred in the morning of April 1, 2471 b.C. (the path of this eclipse was first calculated by Kudlek and Mickxer already in 1971; the eclipse is listed as an event of possible political relevance by Sellers (1992) and first mentioned in direct connection with Shepseskaf by Magli (2013)).
The first point to be examined is, of course, if this date is compatible with the king's regnal years. Chronology of the Old Kingdom (and in particular of the IV dynasty) is still a subject of debate: the length of each single reign is subject to some uncertainty on its own, and the beginning of each regnal period has to be considered within a safety band which in itself is difficult to estimate but cannot be less than, say, ±25 years.
In table 1, four among the most accepted chronologies for the IV dynasty are reported: (1) Baines and Malek (1981), (2) von Beckerath (1997), (3) Shaw ( 2000), (4) Hornung, Krauss and Warburton (2006) (the asterisks signal a short reign attributed to "Bicheris" in (2) and (3) ). In all four chronologies the year 2471 b.C. can be accommodated at the beginning of Shepseskaf's reign. Furthermore, strong hints exist pointing to the planning of the Khufu pyramid – and thus to Khufu's first year - in 2550 b.C.. with the impressive uncertainty of only 10 years; these hints come from the alignment of this pyramid to the cardinal points (Belmonte 2001). If this astronomical dating of Khufu is accepted, then as a matter of fact the "low" and the "high" chronologies of columns (3) and (4) – which are the most unfavourable - are excluded.

A second point of attention is the reliability of calculations of historical solar eclipses.
The mechanics of the alignments of the Sun, the Moon and the Earth can be calculated back in time in a relatively easy way. In other words, we can be sure about when the three celestial bodies have been aligned in the past (recall that the orbital plane of the Moon is tilted about 5° with respect to the Ecliptic, so that solar/lunar eclipses can only occur when the new/full Moon is close to a node). Thus, a total solar eclipse did occur on April 1, 2471 b.C. However, the Earth does not rotate with constant speed, so the problem relies in the identification of the totality band, that is, in establishing where on the Earth the umbra of the Moon was actually projected and so the eclipse was seen as total. The task of researchers working in this fascinating field is thus to estimate the parameter – usually denoted by ΔT - which affects the position and timing of the totality band. This parameter measures the cumulative effect of the rotational change, and can be calibrated using historical sources (see e.g. Hayakawa et al. 2021, 2022; Stephenson et al. 2016, Morrison & Stephenson 2001,2004). The estimate of ΔT has been, in the last 20 years or so, continuously improved and we have today a quite reliable formula (it should be noted, however, that the last registered event we are aware of is of the $8^{th}$ century b.C., and therefore all calculations for eclipses that predate this event are based on extrapolation). This formula is based on the best fit of an impressive amount of data, and expresses ΔT as a parabolic function of the elapsed time (of course, as as for any experimental data, on ΔT there is an uncertainty, say δ, which increases back in time and affects the position of the totality band) (Morrison et al. 2021, Espinak 2006, Gautschy 2012).
For April 1 2471 b.C. the formula gives $\Delta T \approx 16$ h 24, and the part of the calculated path crossing over Egypt is shown in Fig. 3. As we can see, the "capital" Memphis (the likely place of main residence of the Pharaoh, although the concept of capital was not strict in the Old Kingdom) is out of the totality band but very near, by all means within the area in which brilliant stars can actually be seen )[1] . Heliopolis (30 Kms further north) is almost on the verge of totality, which fully involved the delta of the Nile and, in particular, Buto. On that day, the Sun rose already dramatically eclipsed, with the phenomenon reaching totality around 7.59 a.m.. Our star had declination -4° (spring equinox was still to come, on April 11 p. Julian). The eclipsed Sun was between Aries and Taurus, and brilliant stars like the decans Aldebaran and Capella were certainly visible. The Pleiades were so close to the eclipsed Sun that I would venture to suppose that they might have been seen, although their magnitude is low (as an asterism it is between 3.5 and 4). Impressively visible were the two planets Venus and Mercury, in a sort of symmetric configuration along the ecliptic, with the coupled Sun/Moon in the middle (Fig. 4). The eclipse totality lasted a terrifying amount of time: almost seven minutes (eclipses cannot last more than about 7 min 30 sec).

For the sake of clarity, it should be noted that this description holds for ΔT equal or close to its expected value. It is possible to analyse how the uncertainty eventually affects the position and timing of the totality band, and it turns out that the totality coverage of lower Egypt remains essentially the same – with the beginning of totality moving towards sunrise – for the entire interval of values between $\Delta T-\delta$ and ΔT. Instead in the case of positive increment the totality band bends northward, eventually up to the point of living Egypt. Although unlikely, at the present status of knowledge, it is impossible to exclude categorically this last possibility.

## 4. CONCLUSION

The bulk of the chronology of ancient Egypt is based on registered astronomical observations of Lunar dates and of Heliacal risings of Sirius. These observations (leaving aside the problems which occur in analyzing them, for a complete and up-to-date discussion see Belmonte and Lull 2024) actually furnish *astronomical anchors* on which specialists can rely in dating the regnal periods of each Pharaoh. Unfortunately, the first available registrations of Sothic dates come from the Middle Kingdom (for a possible exception see Gautschy et al. 2017), so no Sothic anchor is as yet available for the Old Kingdom. However, astronomy can be of help also in ways which differ from astronomical registrations, one example being the dating of the alignment of the Khufu pyramid already mentioned. The results of the present paper - which arise as well from the interplay between architecture and astronomy, although not in the sense of astronomical alignments - can be considered as giving another, new and quite accurate estimate for the regnal years of an Old Kingdom Pharaoh.

[1] I am grateful to prof. Hirashi Hayakawa for this observation

-

The author reports there are no competing interests to declare.

**REFERENCES**

Assmann, J. (2001). The Search for God in Ancient Egypt. Cornell University Press.
Assmann, J. (2007) *Creation Through Hieroglyphs: The Cosmic Grammatology Of Ancient Egypt* in *Jerusalem Studies in Religion and Culture* 6 Brill, NY
Baines J. and Malek, J. (1981) *The Cultural Atlas of the World: Ancient Egypt*, Oxford Un. Press, Oxford.
Barsanti A, Maspero G. ( 1907) *Fouilles de Zaouiét el-Aryân (1904-1905-1906)* Annales du Service des Antiquités de l'Egypte, 201 - 210
Barta, M. (2005) *Location of Old Kingdom pyramids in Egypt*, Cambridge Arch. Journal 15, 177-191.
Belmonte, J.A. (2001) *On the Orientation of Old Kingdom Egyptian Pyramids*, Archaeoastronomy 26, S1.
Belmonte, J. A., & Lull, J. (2023). Astronomy of Ancient Egypt: A Cultural Perspective. Springer Nature.
Belmonte, J. A., & Magli, G. (2015). Astronomy, architecture, and symbolism: the global project of Sneferu at Dahshur. Journal for the History of Astronomy, 46(2), 173-205.
Bogdanov, I. (2019). The old kingdom evidence on the toponym ẖntj-š "lebanon". Ägypten und Levante/Egypt and the Levant, 29, 125-148.
Breasted, J.H. (1906) *Ancient Records of Egypt,* vol. II, Chicago, 1906,
Brewer, B (1991) *Eclipse.* Earth View, Seattle.
Espenak, F. (2006) Five Millennium Canon of Solar Eclipses: -1999 to +3000. http://eclipse.gsfc.nasa.gov/SEcat5/SEcatalog.html
Gautschy, R. (2012) Sonnenfinsternisse und ihre chronologische Bedeutung: Ein neuer Sonnenfinsterniskanon für Altertumswissenschaftler, Klio 94, Vol. 1, p. 7-17.
Gautschy, R., Habicht, M. E., Galassi, F. M., Rutica, D., Rühli, F. J., & Hannig, R. (2017). A New Astronomically Based Chronological Model for the Egyptian Old Kingdom. Journal of Egyptian History, 10(2), 69-108
Hassan, Selim (1936). Excavations at Gîza II, 1930-1931. Cairo: Faculty of Arts of the Egyptian University & Government Press.
Hayakawa, H., Murata, K., & Sôma, M. (2022). The Variable Earth's Rotation in the 4th–7th Centuries: New ΔT Constraints from Byzantine Eclipse Records. Publications of the Astronomical Society of the Pacific, 134(1039), 094401.
Hayakawa, H., Sôma, M., & Kinsman, J. H. (2021). Analyses of a datable solar eclipse record in Maya Classic period monumental inscriptions. Publications of the Astronomical Society of Japan, 73(6), L31-L36.
Hawass, Z. (1993) *The Great Sphinx at Giza: Date and Function* In *Sesto Congresso Internazionale di Egittologia*. Ed. G.M. Zaccone and T. Ricardi di Netro, 177–195.Turin.

Hamilton, K. (2018) Mastaba el-Faraun, a laymans guide.
Hornung, E., Krauss R. and Warburton D.A. (2006) *Ancient Egyptian Chronology* Leiden.
Kudlek, M and Mickxer, E. (1971) Solar and Lunar Eclipses of the Ancient Near East from 3000 B.C. to 0 with Maps. Alter Orient und Altes Testament ix. Butzon & Bercker Verlag.
Jeffreys, D. (1998) *The Topography of Heliopolis and Memphis: some cognitive aspects,* In *Beitrage zur Kulturgeschichte Ägyptens, Rainer Stadelmann gewidmet* (Mainz) 63-71.
Jéquier, G. (1925). "Le Mastabat-el-Faraoun et le culte funéraire de Shepseskaf". Comptes rendus des séances de l'Académie des Inscriptions et Belles-Lettres Année 69 (4): 251–261.
Lehner, M., (1985) *A contextual approach to the Giza pyramids*, Archiv fur Orientf.. 31, 136-158
Lehner, M. (1999) *The complete pyramids*, Thames and Hudson, London.
Magli, G. (2008) *Akhet Khufu: Archaeo-astronomical Hints at a Common Project of the Two Main Pyramids of Giza, Egypt*. Nexus Network Journal- Architecture and Mathematics 11, 35-50.
Magli, G. (2010) *Topography, astronomy and dynastic history in the alignments of the pyramid fields of the Old Kingdom*; Mediterranean Archaeology and Archaeometry 10, 59-74.
Magli, G. (2013) Architecture, astronomy and sacred landscape in ancient Egypt. Cambridge: Cambridge University Press.
Magli, G. (2016). The Giza 'written' landscape and the double project of King Khufu. Time and Mind, 9(1), 57-74.
Magli, G. (2023). Satellite-Aided Analysis of the Position of the Sun Temples and the Dynastic History of the Vth Egyptian Dynasty. Heritage, 6(11), 7156-7169.
Morrison, L. V., & Stephenson, F. R. (2004). Historical Values of the Earth's Clock Error Δ and the Calculation of Eclipses. Journal for the History of Astronomy, 35(3), 327-336.
Morrison, L. V., Stephenson, F. R., Hohenkerk, C. Y., & Zawilski, M. (2021). Addendum 2020 to 'Measurement of the Earth's rotation: 720 BC to AD 2015'. Proceedings of the Royal Society A, 477(2246), 20200776.
Morrison, L. V., & Stephenson, F. R. (2001). Historical eclipses and the variability of the Earth's rotation. Journal of Geodynamics, 32(1-2), 247-265.
McMurray, W. (2005). Dating the Amarna Period in Egypt: Did a Solar Eclipse Inspire Akhenaten? www.egyptologyforum.org
Reisner, Georges A. (1931). Mycerinus: The Temples of the Third Pyramid at Giza. Cambridge, MA.: Harvard University Press. OCLC 248947316.
Rowe, A. (1940). Newly dentified monuments in the Egyptian Museum showing the deification of the dead together with brief details of similar objects elsewhere. ASAE, 40, 1-67.
Quirke, S. (2001) *The Cult Of Ra* Thames and Hudson, London.
Shaltout, M., Belmonte, J. A., & Fekri, M. (2007). On the orientation of ancient Egyptian temples:(3) Key points at lower Egypt and Siwa Oasis, Part II. Journal for the History of Astronomy, 38(4), 413-442.
Shaw, Ian (2000). *The Oxford History of Ancient Egypt*. Oxford University Press
Sellers, J. (1992) *The Death of Gods in Ancient Egypt* Penguin, London 1992
Stephenson, F. R., Morrison, L. V., & Hohenkerk, C. Y. (2016). Measurement of the Earth's rotation: 720 BC to AD 2015. Proceedings of the Royal Society A: Mathematical, Physical and Engineering Sciences, 472(2196), 20160404.
Verner, M. (2002) *The Pyramids: The Mystery, Culture, and Science of Egypt's Great Monuments* Grove Press, NY
Verner, M. (2006) *Contemporaneous Evidence for the Relative Chronology of Dyns. 4 and 5* In Ancient Egyptian Chronology, 124–143. Edited by Erik Hornung, Rolf Krauss, and David A. Warburton. Leiden/Boston: Brill, 2006.
Von Beckerath, J. (1997) *Chronologie des Pharaonischen Ägypten* Mainz.

**Table 1. Chronology of the IV Dynasty**.

| King | 1 | 2 | 3 | 4 |
|---|---|---|---|---|
| Snefru | 2575-2551 | 2589–2554 | 2543–2510 | 2613–2589 |
| Khufu | 2551-2528 | 2554–2531 | 2509–2483 | 2589–2566 |
| Djedefra | 2528-2520 | 2531–2522 | 2482–2475* | 2566–2558 |
| Khafra | 2520-2494 | 2522–2496 * | 2472–2448 | 2558–2532 |
| Menkaura | 2494-2472 | 2489–2461 | 2447–2442 | 2532–2503 |
| Shepseskaf | 2472-2465 | 2461–2456 | 2442–2436 | 2503–2498 |

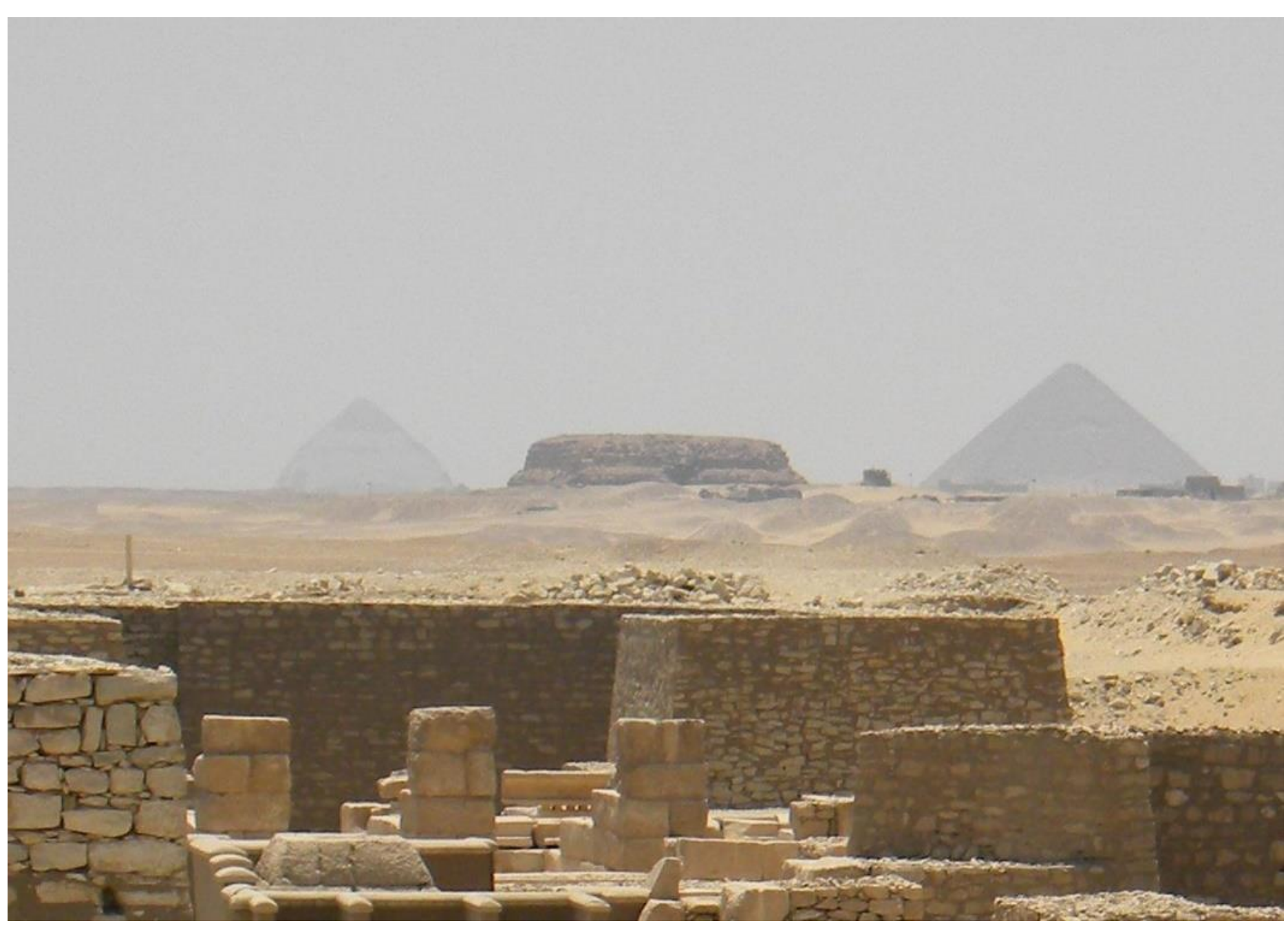

Figure 1

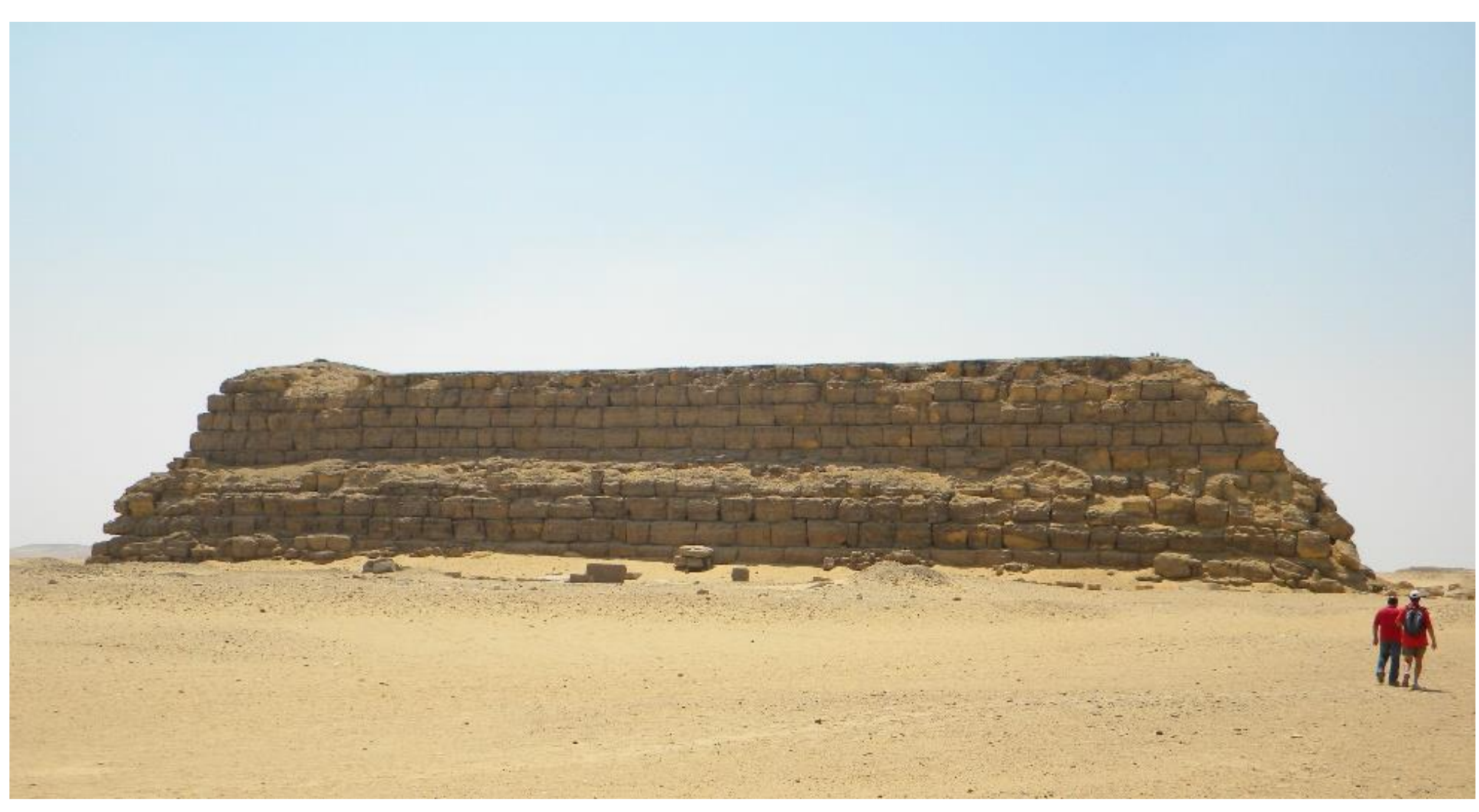

Figure 2

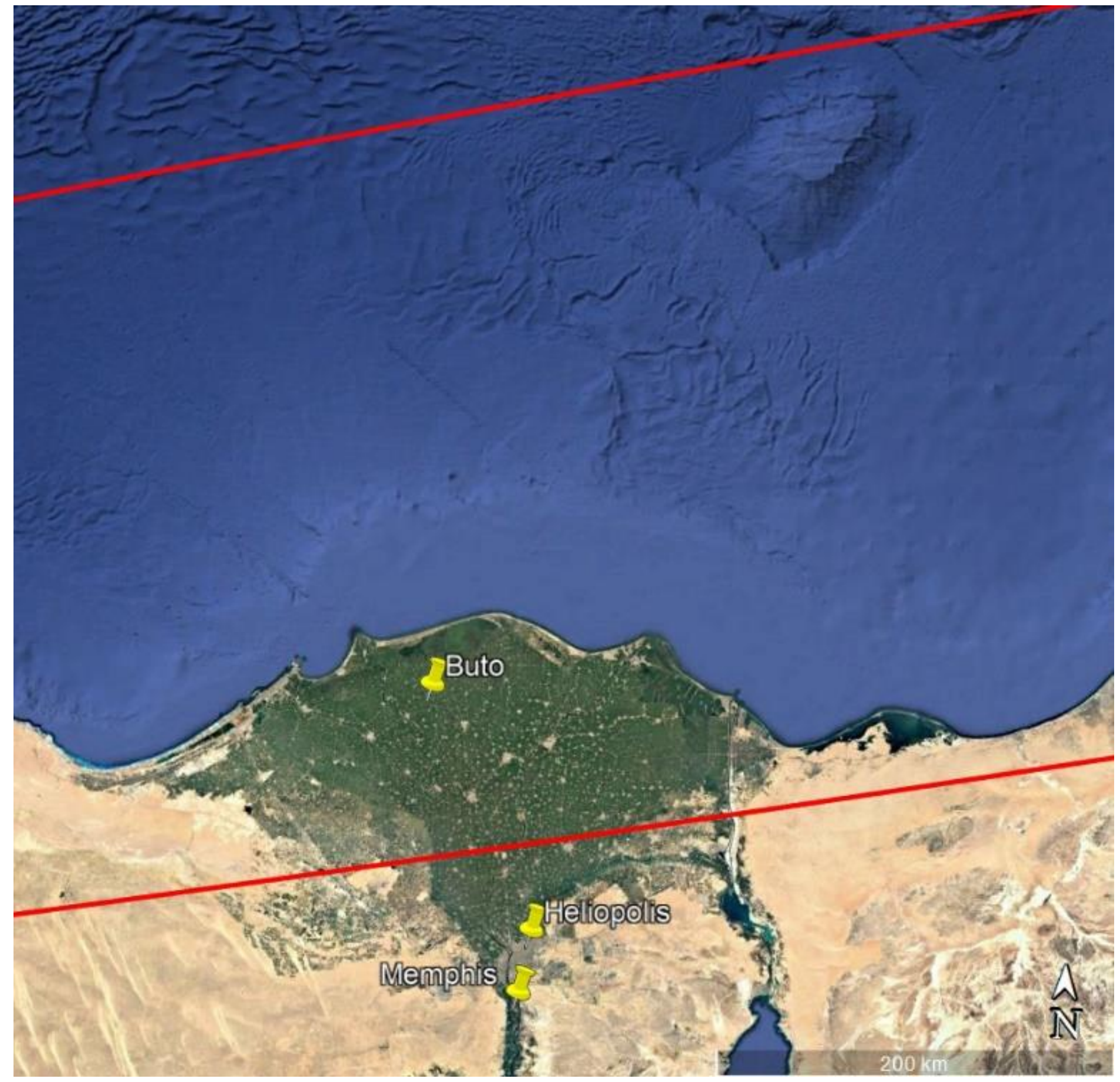

Figure 3

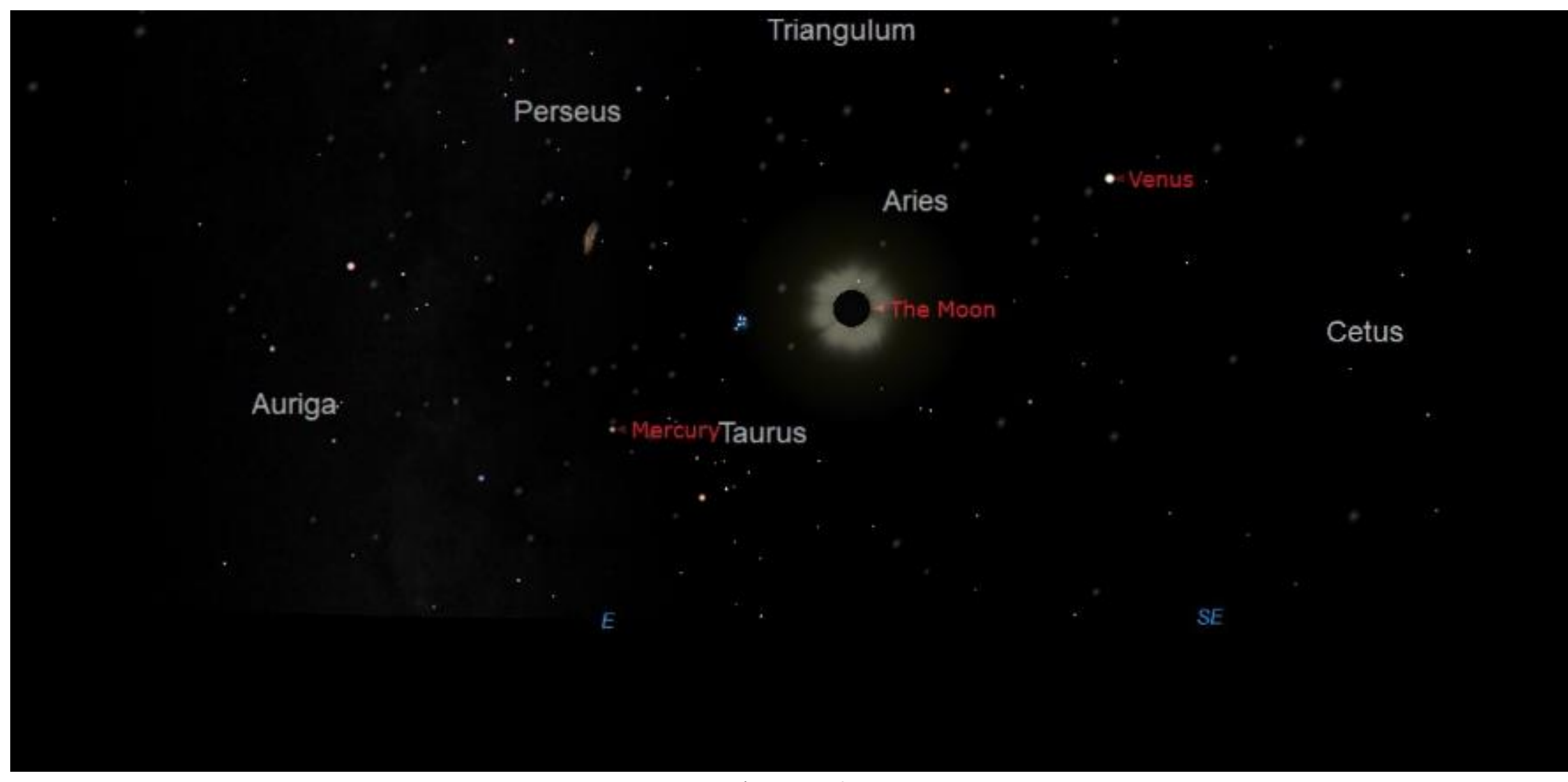

Figure 4

**Figure 1.** View of the southern horizon from Saqqara south. Image: author.

**Figure 2.** The Shepseskaf tomb, view from the east. Image: author.

**Figure 3.** The April 1 2471 b.C. eclipse totality band over Egypt (Image courtesy Google Earth, data from Espenak 2006 and Gautschy 2012).

**Figure 4.** The sky during totality. Image: author elaboration with Starry Night Pro.